\documentclass[12pt]{iopart}
\usepackage[numbers]{natbib}
\usepackage{graphicx}

\begin{document}

\title{Stability of the Antiferromagnetic State in the Electron Doped Iridates}

\author{Sayantika Bhowal, Jamshid Moradi Kurdestany and Sashi Satpathy}

\address {Department of Physics \& Astronomy, University of Missouri, Columbia, MO 65211, USA}

\begin{abstract}
Iridates such as Sr$_2$IrO$_4$ are of considerable interest owing to the formation of the Mott insulating state driven by a large spin-orbit coupling.  However, in contrast to the expectation from the Nagaoka Theorem that a single doped hole or electron destroys the anti-ferromagnetic (AFM) state of the half-filled Hubbard model in the large U limit, the anti-ferromagnetism persists in the doped Iridates  for a large dopant concentration beyond half-filling. 
With a tight-binding description of the relevant J$_{\rm eff}$ = 1/2 states by the third-neighbor  ($t_1$, $t_2$, $t_3$, $U$) Hubbard model on the square lattice, we examine the stability of the AFM state to the formation of a spin spiral state in the strong coupling limit. 
The third-neighbor interaction $t_3$  is important for the description of the Fermi surface of the electron doped system. A phase diagram in the parameter space is obtained for the regions of stability of the AFM state. Our results qualitatively explain the robustness of the AFM state in the electron doped iridate (such as Sr$_{2-x}$La$_x$IrO$_4$),  observed in many experiments, where the AFM state continues  to be stable until a critical dopant concentration.
\end{abstract}
\maketitle
\section{INTRODUCTION}

Recently iridates are being vigorously investigated both theoretically as well as experimentally due to the various exotic phases hosted by them that includes the J$_{\rm eff}$ = 1/2 Mott insulating state~\cite{Kim}, quantum spin liquid state~\cite{BYIO,BZIO}, Weyl semi-metal phase~\cite{WS}, etc. The importance of the spin-orbit coupling (SOC) in the iridates was first realized in $d^5$ iridate Sr$_2$IrO$_4$ (SIO) ~\cite{Kim}, where the anti-ferromagnetic (AFM) insulating state is essentially an outcome of the interplay between the spin-orbit coupled pseudo-spin J$_{\rm eff}$ = 1/2 states and onsite Coulomb repulsion ($U$). The quasi two-dimensional layered structure of the iridate, isostructural to high-T$_c$ cuprate La$_2$CuO$_4$, has led to the interest to study the electron doped system in  search of unconventional superconductivity. This results in a large number of experimental work~\cite{Ge,Korneta,Brouet,Chen} on the electron-doped iridate in the past decade.

Recent experiments on the electron-doped iridate~\cite{Chen} reveal that the long range AFM interaction of the undoped (half-filled) system persists up to a doping concentration $x \simeq 0.04$ in Sr$_{2-x}$La$_x$IrO$_4$, beyond which short-range AFM order survives up to the highest doping level of La indicating the presence of robust AFM interaction in the system. This is however not obvious at a first glance as according to the Nagaoka theorem~\cite{Nagaoka}, very small carrier doping can destabilize the AFM state in a nearest neighbor (NN) model in the presence of strong Coulomb repulsion. 
In addition, a recent experiment has found a spin-density-wave (SDW) ground state when the dopant concentration exceeds
$x \simeq 0.08$ \cite{ChenNature}.
 
 In view of the above mentioned scenario, we have examined the stability of the AFM state  with respect to the spiral instability (formation of SDW) using the third-neighbor ($t_1$, $t_2$, $t_3$, $U$) Hubbard model on a square lattice constructed in the pseudo-spin J$_{\rm eff}$ = 1/2 basis relevant to describe the spin-orbit coupled states of the iridate. 
 Our stability analysis is performed in the strong-coupling limit $U/W \rightarrow \infty$, where $W$ is the band width.
 The stability of the doped Hubbard model has been studied earlier \cite{Avinash} in connection with the high-T$_{\rm c}$ cuprates. 
 %
 Our stability analysis is performed within the 
 Hartree-Fock (HF) approximation and the second-order perturbation theory for the $M$-electron pocket, which is relevant for the electron doped SIO. 
  Furthermore, unlike the cuprates, the third-neighbor interaction $t_3$ is not only important for the location of the electron pocket at the 
M point in SIO, but  is also  found to play a crucial role in determining  the stability of the AFM state. The phase diagram in the parameter space $t_3$-$U$ and $x$-$U$ are obtained to describe the regions of stability, thereby providing an understanding of the observed robustness of the AFM state in the electron-doped iridate.
The results show that the AFM state is stable for small doping concentration in qualitative agreement with the experimental findings.
 
 The rest of the paper is organized as follows. In Section \ref{sec2}, we discuss the model Hamiltonian, obtain the parameters by fitting  to the density-functional bands, and discuss the band structure features relevant to the electron-doped case. This is followed in Section \ref{sec3} by the stability analysis of the AFM state, considering  the role of both the intra and inter-band transitions.
 We then show the phase diagram in the parameter space obtained from the instability condition and discuss its connection
 with the experimental results. The results are summarized in Section \ref{sec4}. 

\section{MODEL HAMILTONIAN}    \label{sec2}   

\begin{figure}[h]
 \centerline{\includegraphics[width=150pt]{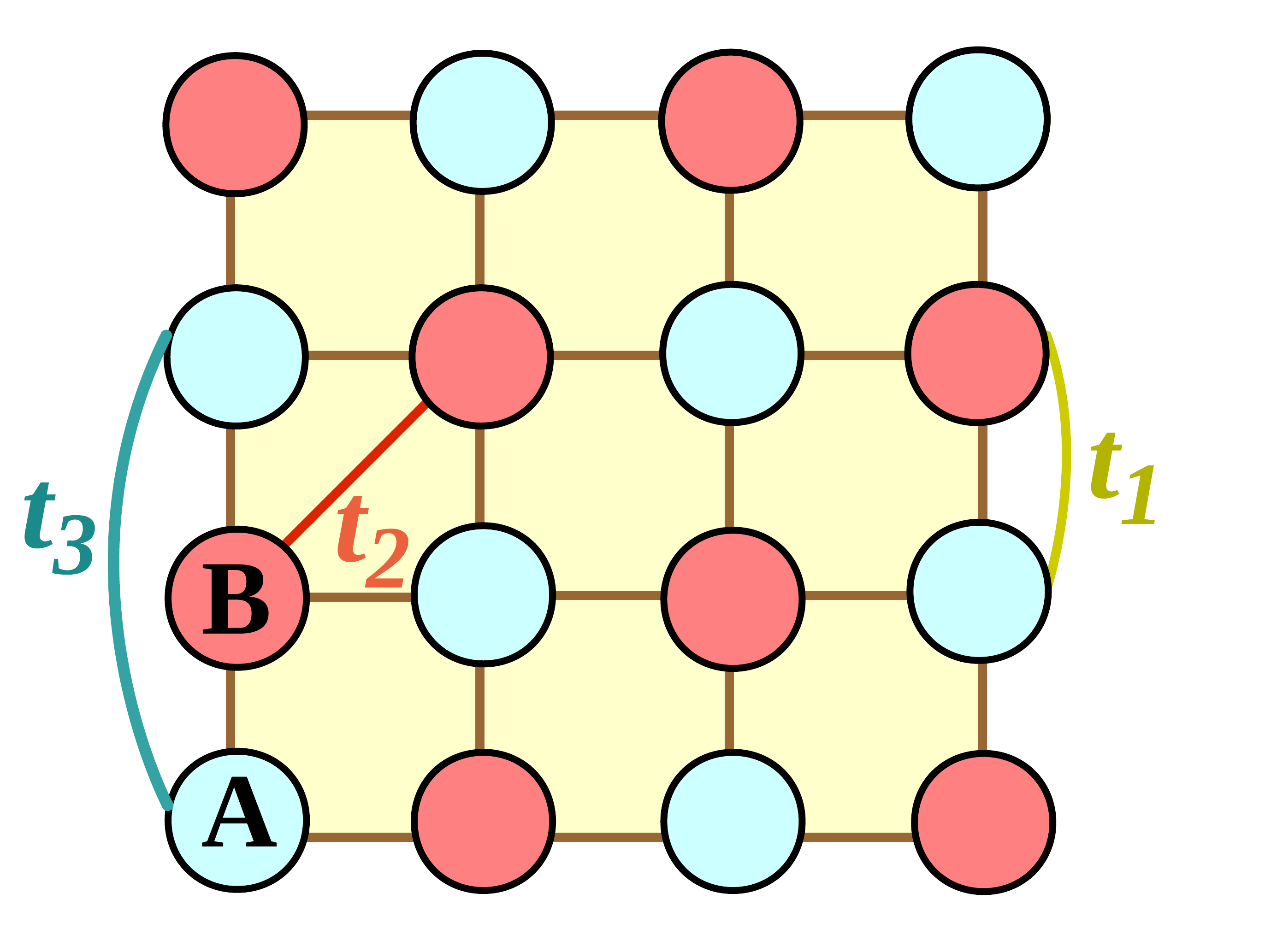}}
  \caption{ (color online) The square lattice, relevant for Sr$_2$IrO$_4$ considered in the paper, with  two different sublattices, A and B, marked in different colors denoting Ir atoms with two different pseudo-spins. The NN, next NN, and third NN hoppings are also indicated.}
\label{fig1}
\end{figure}

We consider the third-neighbor Hubbard model on a square lattice, relevant for the layered structure of the Ir atoms in SIO,
\begin{eqnarray}
\label{Hubbardmodel}
{\cal H} & = &\sum_{ij\sigma} t_{ij}  \ (c_{i\sigma}^{\dagger} c_{j\sigma} + H. c.) 
+ U\sum_i { n}_{i\uparrow}{ n}_{i\downarrow}
\end{eqnarray}
where $c_{i\sigma}^{\dagger}$ ($c_{i\sigma}$) is the electron creation (annihilation) operator at site $i$ with spin $\sigma$, 
$n_{i\sigma} = c^{\dagger}_{i\sigma}c_{i\sigma}$ is the number operator,
 $U$ is the Coulomb repulsion, and the hopping integrals $t_{ij}$ 
 are kept up to the third NN
  $( t_1, t_2,  t_3)$, a minimal model to describe the J$_{\rm eff}$ = 1/2 upper and lower Hubbard bands of SIO. 
  Note that we have neglected the hopping between the Ir layers normal to the plane, because of the large separation between the layers. This is further justified from density-functional band structure which shows negligible dispersion for momentum along the direction normal to the plane \cite{Churna}.
  The hopping integrals are 
  indicated in Fig. \ref{fig1} and the spin-orbit coupled pseudo-spin basis states 
 are defined as

\begin{eqnarray}
  |e_1 \rangle & \equiv & |\frac{1}{2},-\frac{1}{2} \rangle  = \frac{1}{\sqrt{3}}(|xy \uparrow\rangle + |yz \downarrow\rangle + i |xz \downarrow\rangle) \nonumber \\ 
 |e_2 \rangle & \equiv & |\frac{1}{2}, \frac{1}{2} \rangle  =  \frac{1}{\sqrt{3}}(|yz \uparrow\rangle -i |xz \uparrow\rangle - i |xy \downarrow\rangle),
 \label{jstate}
\end{eqnarray}  
so that $c_{i\sigma}^{\dagger}$ creates an electron in the state  $ |e_\sigma \rangle$ at site $i$.
The 1NN tight-binding hopping integral  $t_1$ connecting orbitals on two different sublattices  becomes a complex number,
which however can be made real for the sake of convenience by a gauge transformation \cite{Mohapatra} that adds a 
 suitable phase factor to the definitions of the orbitals  in Eq.  \ref{jstate}.
 We assume this has been done and all our hopping parameters are thus real. Note that hoppings are allowed only between orbitals of the same pseudospin.
%

The tight-binding Hamiltonian is easily obtained with the HF approximation. 
We consider the AFM state.
In the Bloch function basis set of the pseudo-spin orbitals   $(b^\dagger_{A\uparrow}, b^\dagger_{B\uparrow}, b^\dagger_{A\downarrow}, b^\dagger_{B\downarrow})$,  
the Hamiltonian becomes block diagonal, since hopping does not change the pseudo spin. 
The Bloch functions, expressed in terms of the field operators, are simply
$b^\dagger_{\alpha\sigma} (k) = N^{-1/2} \sum^\prime_i \ \exp (i k \cdot r_i) \ c^\dagger_{i\sigma} $, where the prime 
indicates summation over all atoms in the sublattice $\alpha$.
Thus we have
\begin{eqnarray}\label{Hk1}
{H}_{HF} (k) = 
\left[ 
{\begin{array}{*{20}c}
 { \cal H}^{\sigma = +1}  &    \   0      \\
          0       &     \    { \cal H}^{\sigma = -1}  \\
 \end{array} }  \right], 
 \end{eqnarray}
where 
\begin{eqnarray}
\label{Hk2}
{ \cal H}^{\sigma}(k) = 
\left[ 
{\begin{array}{*{20}c}
-\sigma \Delta + h_{11}  &    \     h_{12}    \\
  h_{12} &          \ \ \              \sigma \Delta + h_{11}  
 \end{array} }  \right] 
 \end{eqnarray}    
 are the 2$\times$2 matrices having identical eigenvalues and eigenvectors.
Here,  $h_{11} = \varepsilon^{(2)}_k+\varepsilon^{(3)}_k$, $h_{12} = \varepsilon^{(1)}_k$, $\sigma = \uparrow, \downarrow$,   
$\Delta = U/2$ is the staggered field, 
and the
expressions for the Bloch sums are given by $\varepsilon^{(1)}_k = 2t_1(\cos k_x+\cos k_y)$, $\varepsilon^{(2)}_k = 4t_2 \cos k_x \cos k_y$, and
$\varepsilon^{(3)}_k = 2t_3(\cos 2k_x +\cos 2k_y)$. 
Note that we have approximated the staggered field to be 
$\Delta = U (n_\uparrow - n_\downarrow) / 2 =  U (1-x)/2 \approx U/2$, 
since we are working close to half filling 
(doping concentration $x$ is small), and the neglected term does not contribute to the transverse response to the first order in $x$.
%
\begin{figure}[t] \label{fig2}
  \centerline{
\includegraphics[width=200pt]{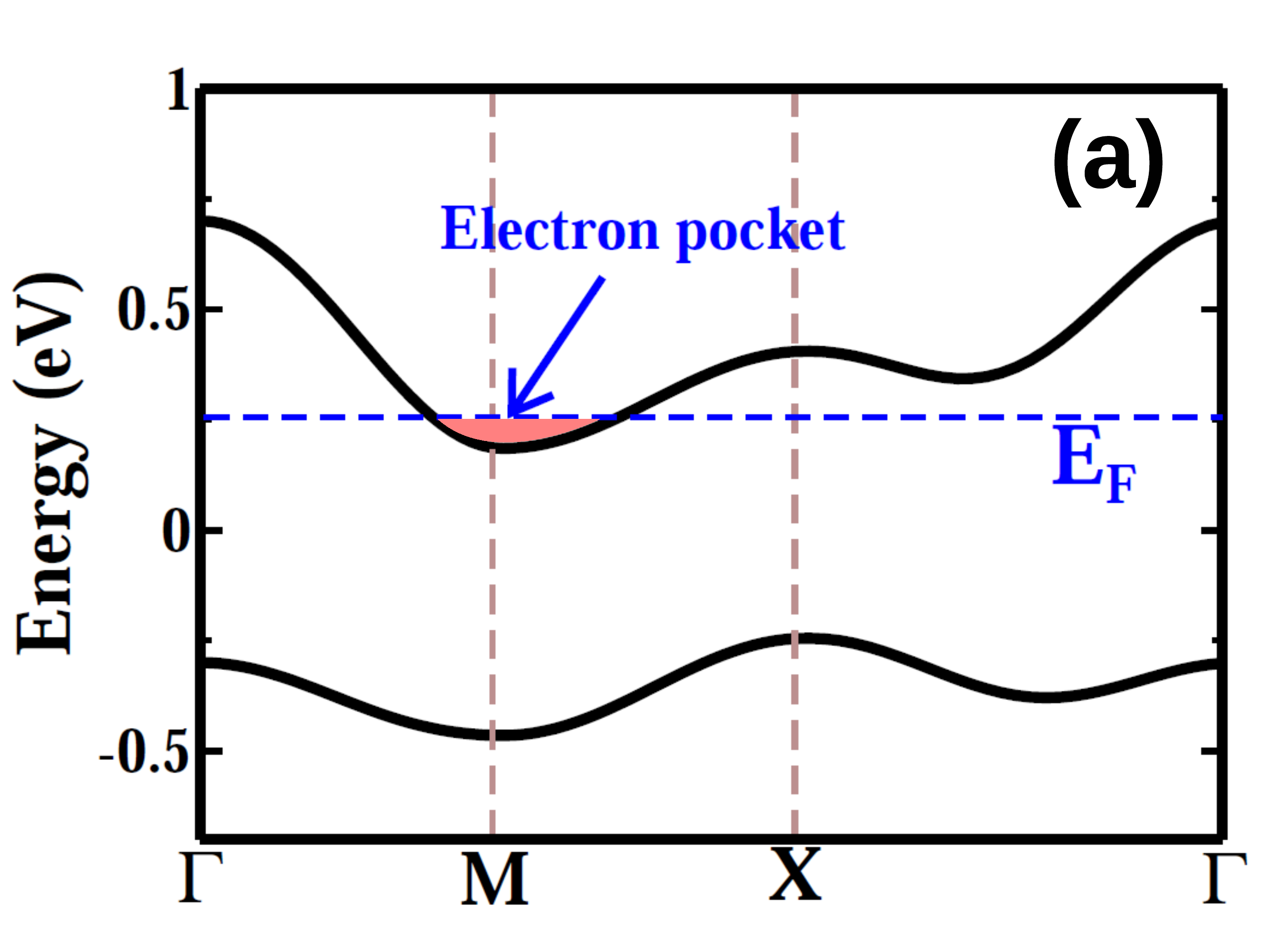}
\includegraphics[width=200pt]{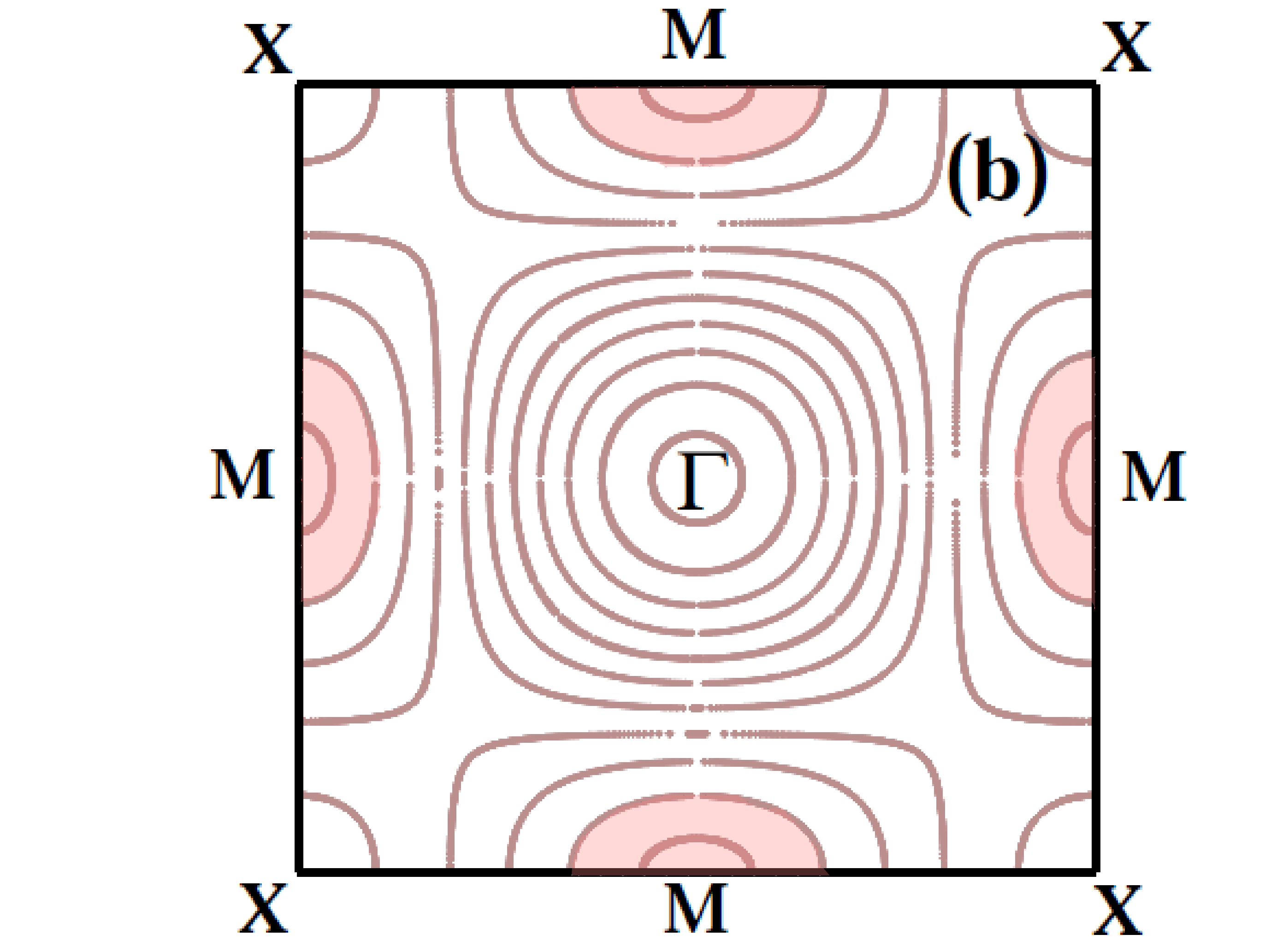}        
}
\caption{ (color online) (a) Tight-binding band structure of SIO with the parameters
 (in eV): 
  $t_1$ = -0.095, $t_2$ = 0.015, $t_3$ = 0.035, and $U$ = 0.65,  obtained by fitting to the density-functional bands \cite{Churna}.  The doped electrons form an electron pocket at the M point.
    The  symmetry points in the BZ are:
$\Gamma$ = (0,0), X = $\pi/a(1,0)$ and M = $\pi/(2a) (1,1)$, where $a$ 
 is the NN distance, with the
coordinate axes chosen along the NN directions, and $ a = 1$ throughout the paper.
(b) The energy contours for the conduction band relevant for electron doping. The contour values are in steps of 0.05 eV with the energy of the $M$ point taken as zero. For small dopant concentration, the doped electrons form elliptical Fermi pockets
around the $M$ points as shown by the shaded region.
  }
\end{figure}
The quasi-particle energies of the above HF Hamiltonian are
\begin{equation}\label{eigenval}
{\varepsilon}^{\pm}_{k \sigma} = \varepsilon^{(2)}_k + \varepsilon^{(3)}_k \pm {\sqrt{\Delta^{2}+(\varepsilon^{(1)}_k)^{2}}},
\end{equation}
with the $+/-$ branches forming the conduction and the valence bands, respectively.  
The corresponding wave functions are
\begin{eqnarray}\label{ef1}
 |\Psi_{k \uparrow}^{c, v} \rangle  = 
\left( 
{\begin{array}{*{20}c}
  \psi_{k \uparrow}^{c, v}    \\   \phi_{k \uparrow}^{c, v}  \\ 0 \\ 0 \\
 \end{array} }  \right),
 \hspace{5mm}
 |\Psi_{k \downarrow}^{c, v} \rangle  = 
\left( 
{\begin{array}{*{20}c}
 0 \\ 0 \\  \psi_{k \downarrow}^{c, v}    \\   \phi_{k \downarrow}^{c, v}  \\
 \end{array} }  \right), 
 \end{eqnarray} 
where
$\psi_{k\sigma}$, $\phi_{k\sigma}$ are the quasi-particle amplitudes for the sites A and B, respectively.  
These are given by
\begin{eqnarray}\label{ef2}
&& (\psi^{v}_{k\uparrow})^2 = (\phi^{v}_{k\downarrow})^2 = (\psi^{c}_{k\downarrow})^2 = (\phi^{c}_{k\uparrow})^2 = \frac{1}{2}\left( 1 + \Delta/\sqrt{\Delta^{2}+(\varepsilon^{(1)}_k)^{2}} \right), \nonumber \\
\hspace {- 2cm} 
{\rm and}  &&(\psi^{c}_{k\uparrow})^2 = (\phi^{c}_{k\downarrow})^2 = (\psi^{v}_{k\downarrow})^2 = (\phi^{v}_{k\uparrow})^2 = \frac{1}{2}\left( 1 - \Delta/\sqrt{{\Delta^{2}+(\varepsilon^{(1)}_k)^2}} \right).
\end{eqnarray} 
 
The tight-binding (TB) parameters as obtained for SIO by fitting to the J$_{\rm eff}$ = 1/2 density-functional bands are (in eV): $t_1$ = -0.095, $t_2$ = 0.015, $t_3$ = 0.035, $t_4 =0.01$, and $U$ = 0.65. 
However, the 4NN hopping $t_4$ being small, we ignore it in our analysis in this paper.
This minimal model explains the band structure of SIO rather well with the electron (hole) pocket at the $M$ ($X$) point of the Brillouin zone (BZ) as shown in Fig.~\ref{fig2} (a). The energy contours for the conduction band around the $M$ point is shown in Fig.~\ref{fig2} (b). 

In this work, we will restrict to  electron doping, and the occurrence of the conduction minimum at the $M$ point is an important feature of the density functional band structure that must be retained in the TB result. For this to happen, the TB parameters must satisfy certain conditions, 
which we proceed to find now.
From the energy expression Eq.   \ref{eigenval}, we find that the $M$ electron pocket is elliptical with the minor/major axes in the diagonal directions $k_1, k_2 =  (k_x \pm k_y)/ \sqrt 2$. 
Expanding the conduction band energy around the $M = -\pi/2 (1,1)$ point, we find for small momentum, the result
\begin{equation}
\varepsilon^+_{k\sigma} = \varepsilon_M +  k_1^2 (2t_2 + 4 t_3 + 4 t_1^2/ \Delta) + k_2^2 (4 t_3 - 2 t_2 ),
\end{equation}
which is an ellipse, and here $\varepsilon_M = \Delta - 4 t_3$.  
The conduction bottom at the $X$ point is given by $\varepsilon_X = 4(t_3 - t_2 ) + \Delta$.
It is easy to see from the above equation  that for the $M$ point to have a positive curvature (local minimum), we must have the condition 
\begin{equation} \label{condition}
t_3 > t_2/2 > -t_1^2/U.     \hspace{30mm}  {\rm (condition \  on \  the \   TB \   parameters)}
\end{equation}
The first inequality automatically makes the $M$ conduction minimum below $X$, and therefore Eq. (\ref{condition}) is the only 
necessary condition for the proper description of the  
conduction band structure of SIO. 
The eccentricity of the ellipse $ e  \equiv (1 - \eta^2)^{1/2} \approx 0.78$, where the axis ratio 
$\eta \equiv \beta / \alpha = [ (2 t_3 - t_2) /  (t_2 + 2 t_3 + 4 t_1^2 / U)   ]^ {1/2} \approx 0.62$ for SIO with the approximation $\Delta \approx U/2$.

\section{STABILITY OF THE ANTIFERROMAGNETIC STATE} \label{sec3} 

We consider the stability of the AFM state in the strong-coupling limit, which is true for parameters relevant for SIO. 
This is done by applying a small transverse perturbing field $\Delta_\perp^i = - U  \langle c^\dagger_{i\uparrow} c_{i\downarrow} \rangle$ on the electrons (perturbing potential
$ V_i = - U  \langle c^\dagger_{i\uparrow} c_{i\downarrow} \rangle c^\dagger_{i\downarrow} c_{i\uparrow} + h. c. $  in the HF Hamiltonian),
by making the order parameter $f_i = \langle c^\dagger_{i\uparrow} c_{i\downarrow} \rangle$ non-zero. 
For the AFM  state, the value of $\langle c^\dagger_{i\uparrow} c_{i\downarrow} \rangle$ is zero as there is no spin mixture in the single-particle states.  
The idea is to apply the  perturbing field $\Delta_\perp^i = - U f_i$, compute its effect on the electron states, obtain the resulting field $\tilde \Delta_\perp^j = - U \tilde f_j$,   and see if it builds up or diminishes. 
It is convenient to express the effect on the transverse order parameter $f$ via the response matrix $\chi$,
which in
 the real-space  representation is written as 
$
\tilde f_j = \sum_{k} \chi_{jk} f_k.
$
%
For stability, the maximum eigenvalue of the response matrix should be less than one.

The response matrix can be computed using perturbation theory.
We denote  the eigenstates by ($ E_{l \sigma},  \ | l \sigma\rangle)$, where $\sigma$ is the spin index and $l$ is the combined Bloch and the band index,
and the wave function amplitudes by $\xi^i_{l \sigma}$, 
so that $d^\dagger_{l \sigma} = \sum_i \xi^i_{l \sigma} \ c^\dagger_{i\sigma}$ creates the band states.
 The application of the transverse perturbation $\Delta^i_{\perp}$ will generate a small amplitude of opposite spin  
  at the  site $i$. The first-order correction to the state $| l\sigma\rangle$ is thus given by
\begin{equation}
\label{es}
 \delta | l\sigma\rangle = \sum_m\sum_{\sigma^{\prime}\neq \sigma} \frac{| m\sigma^{\prime}\rangle\langle m\sigma^{\prime}| \Delta^i_{\perp} | l\sigma\rangle}{E_{l\sigma}-E_{m\sigma^\prime}}.
\end{equation}
The resulting transverse order parameter $\tilde f_i$ can be obtained by computing 
the expectation value $  \langle c_{i \sigma}^\dagger c_{i\sigma^\prime}\rangle$ for the perturbed state,
which leads to the equation $ \tilde f_i=  \sum_{j} \chi_{ij} f_j$, where
\begin{equation}
\label{instability}
\chi_{ij} =  U \sum^{E_m>E_F}_{E_l < E_F}\left( \frac{\xi^i_{l\uparrow}\xi^i_{m\downarrow}\xi^j_{m\downarrow}\xi^{j} _{l\uparrow}}{E_{m\downarrow}-E_{l\uparrow}}
+ \frac{\xi^i_{l\downarrow}\xi^i_{m\uparrow}\xi^j_{m\uparrow}\xi^j_{l\downarrow}}{E_{m\uparrow}-E_{l\downarrow}}\right).
\end{equation}
This is the central quantity to be evaluated in the stability analysis for the AFM state. 
In Eq.  (\ref{instability}), the states designated as $m$ and $l$ are, respectively, above and below the Fermi energy $E_F$, and the $N \times N$ matrix $\chi_{ij}$ is the  response matrix connecting the applied transverse field at site $j$ to the resulting field at site $i$. If the largest eigenvalue of $\chi$ is greater than one, the effect is greater than the cause leading to the instability of the AFM state.

There is one more point to make before we proceed to study the stability condition.
Instead of the real space, we will work in the momentum space and examine  the instability of the AFM state with respect to the formation
of the spiral SDW state. 
The spiral perturbing field we consider is $ (f_A, f_B) \exp \ (i   q \cdot   R_i)$,
where  
  $  q \equiv (q_x, q_y)$  is the spiral wave vector, $  R_i$ are the cell positions, and 
  $f_A= \langle c^\dagger_{A\uparrow} c_{A\downarrow} \rangle$ and $f_B= \langle c^\dagger_{B\uparrow} c_{B\downarrow} \rangle$ 
  are the order-parameter amplitudes on the two sublattice atoms in the unit cell. 
  This is essentially a SDW fluctuation of momentum $q$ on top of the AFM state.
  
  The effect of this SDW transverse field is  described by the response matrix $\chi (q)$
\begin{eqnarray}
\left( 
{\begin{array}{*{20}c}
 \tilde f_A       \\
\tilde f_B  \\
\end{array} }  \right)
=
\left( 
{\begin{array}{*{20}c}
  \chi{(q)}_{AA} &    \      \chi{(q)}_{AB}    \\
   \chi{(q)}_{BA} &          \ \ \               \chi{(q)}_{BB}  \\
\end{array} }  \right) 
\left( 
{\begin{array}{*{20}c}
f_A       \\ 
 f_B  \\
\end{array} }  \right)
, 
\end{eqnarray}
whose maximum eigenvalue must be less than one for every $q$ for the stability of the AFM phase.
Since the two sublattices are equivalent, symmetry dictates that
 $ \chi{(q)}_{AA} = \chi{(q)}_{BB}$ and $ \chi{(q)}_{AB} = \chi{(q)}_{BA}$. 
 Of the two eigenvalues of $ \chi{(q)}$,  
 \begin{equation} \label{lambda-stability}
 \lambda^{\rm max}_{\perp}(q) = \chi{(q)}_{AA} - \chi{(q)}_{AB} < 1 \ \ \ \ \ \  \ \hspace {30mm}  {\rm (stability \ condition)}
 \end{equation}
  turns out to be larger due to the sign of
 $\chi{(q)}_{AB}$. 
 This $\lambda^{\rm max}_{\perp}(q)$  should be less than one for all $q$  for the stability of the AFM state as indicated in Eq.  (\ref{lambda-stability}).  

\begin{figure}[h]
  \centerline{\includegraphics[width=250pt]{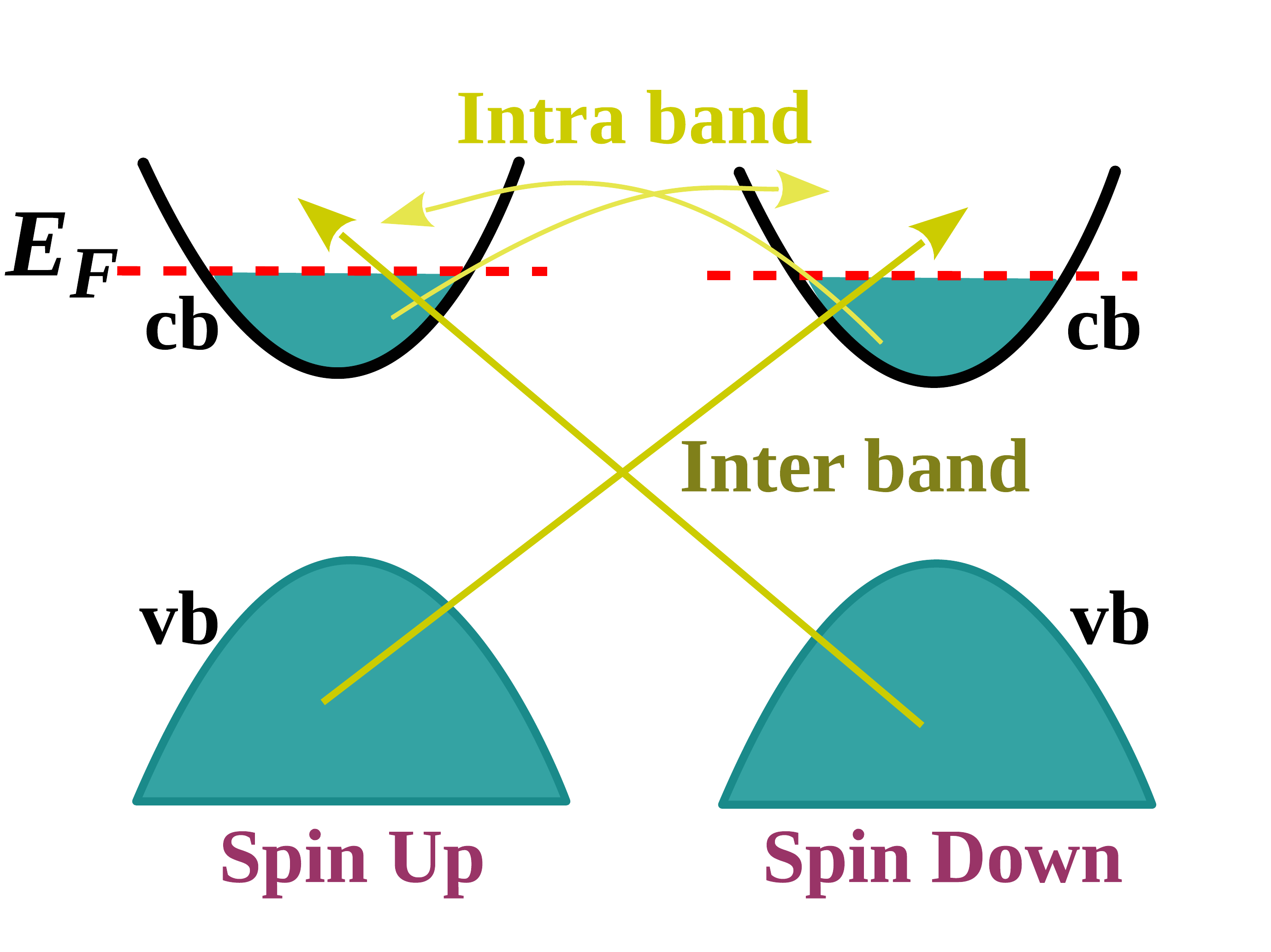}}
  \caption{ (color online) Intra- and inter-band processes in response to a  transverse field  
  $V  \ \langle c_{i\uparrow}^\dagger c_{i\downarrow} +  c_{i\downarrow}^\dagger c_{i\uparrow} \rangle $ that
  seeks to destabilize the AFM state.
  The transverse field mixes the occupied states of one spin with the unoccupied states of opposite spin. 
 }
\label{fig3}
\end{figure}

At exactly half-filled condition (no doping), the valence band is completely filled, while the conduction band is empty. 
In such a situation, only the inter-band transitions between the valence and conduction bands will contribute to the response, while for the 
doped case, both intra and inter-band transitions will contribute (Fig. \ref{fig3}). Note also that for the spiral field,
only the initial and final states in the transition process differing by momentum $q$
 will contribute to the response.
This restricts the summation in the response expression, Eq.  \ref{instability}, which we now proceed to evaluate.

\subsection{Intra-band Contribution} 

We first calculate the intra-band contribution to the response matrix. 
As mentioned earlier, doping induces an electron pocket in the conduction band around the $M$-point with $(k_{x},k_{y}) = \pi \ (\pm1/2,\pm1/2)$ in SIO. Hence, in the intra-band processes, transitions occur from the filled states around the $M$-point to the empty states within the conduction band, separated by the spiral momentum $q$. For this case, Eq.  (\ref{instability})  becomes
\begin{equation} 
[ \chi_{AA}{(q)}]^{\rm intra} = U  \sum_k^{\prime}\left( \frac{(\psi^{c}_{k\uparrow})^2(\psi^{c} _{k+q\downarrow})^2}{\varepsilon^{c}_{k+q\downarrow} - \varepsilon^{c}_{k\uparrow}}
+ \frac{(\psi^{c} _{k\downarrow})^2(\psi^{c} _{k+q\uparrow})^2}{\varepsilon^{c}_{k+q\uparrow} - \varepsilon^{c}_{k\downarrow}}\right)  ,
\end{equation}
where the prime in the summation indicates that the $k$-states are below the Fermi energy,
while $k+q$ states are above,  {\it i. e.}, $k < k_F$ and $k+q > k_F$. 
Substituting the values of the quasi-particle amplitudes and energies from Eqs. (\ref{eigenval}) and (\ref{ef2}), we get the result
\begin{equation}
\hspace{-15mm}
[ \chi_{AA}{(q)}]^{\rm intra} =     U \sum^{\prime}_k \frac
{(4\Delta^2)^{-1} \times   [(\varepsilon^{(1)}_{k})^2+(\varepsilon^{(1)}_{k+q})^2]}
{(\varepsilon^{(2)}_{k+q}-\varepsilon^{(2)}_{k}+\varepsilon^{(3)}_{k+q}-\varepsilon^{(3)}_{k})+((\varepsilon^{(1)}_{k+q})^2-(\varepsilon^{(1)}_{k})^2)/   (2\Delta)   },
\end{equation}
where the strong-coupling limit  $ |t_i / \Delta | \ll  1$ has been assumed. Similarly, for the off-diagonal terms of the response matrix, we find
 \begin{eqnarray} 
&&[ \chi_{AB}{(q)}]^{\rm intra} =  U   \sum_k^{\prime}
\left( \frac{\psi^{c}_{k\uparrow}\phi^{c}_{k\uparrow}\psi^{c} _{k+q\downarrow}\phi^{c} _{k+q\downarrow}}{\varepsilon^{c}_{k+q\downarrow}-\varepsilon^{c}_{k\uparrow}}
+ \frac{\psi^{c} _{k+q\uparrow}\phi^{c} _{k+q\uparrow}\psi^{c} _{k\downarrow}\phi^{c} _{k\downarrow}}{\varepsilon^{c}_{k+q\uparrow}-\varepsilon^{c}_{k\downarrow}}\right)   \nonumber \\
 &=& U   \sum^{\prime}_k   \frac
 {2\varepsilon^{(1)}_{k}\varepsilon^{(1)}_{k+q}/  (4\Delta^2)}
 {(\varepsilon^{(2)}_{k+q}-\varepsilon^{(2)}_{k}+\varepsilon^{(3)}_{k+q}-\varepsilon^{(3)}_{k})+((\varepsilon^{(1)}_{k+q})^2-(\varepsilon^{(1)}_{k})^2)/   (2\Delta)  } 
\end{eqnarray}
Thus the intra-band contribution to the maximum eigenvalue is 
\begin{eqnarray}\label{lambdamax1}
&&[\lambda^{\rm max}_{\perp}(q)]^{\rm intra}      =    [ \chi_{AA}{(q)}]^{\rm intra}-[ \chi_{AB}{(q)}]^{\rm intra} \nonumber \\
 &=&   U  \sum_{k}^{\prime}\frac{(\varepsilon^{(1)}_{k}-\varepsilon^{(1)}_{k+q})^2/ (4\Delta^2) }{(\varepsilon^{(2)}_{k+q}-\varepsilon^{(2)}_{k}+\varepsilon^{(3)}_{k+q}-\varepsilon^{(3)}_{k})+({(\varepsilon^{(1)}_{k+q})^2-(\varepsilon^{(1)}_{k})^2})/   (2\Delta)  }
\end{eqnarray}

\begin{figure}[h]
  \centerline{\includegraphics[width=250pt]{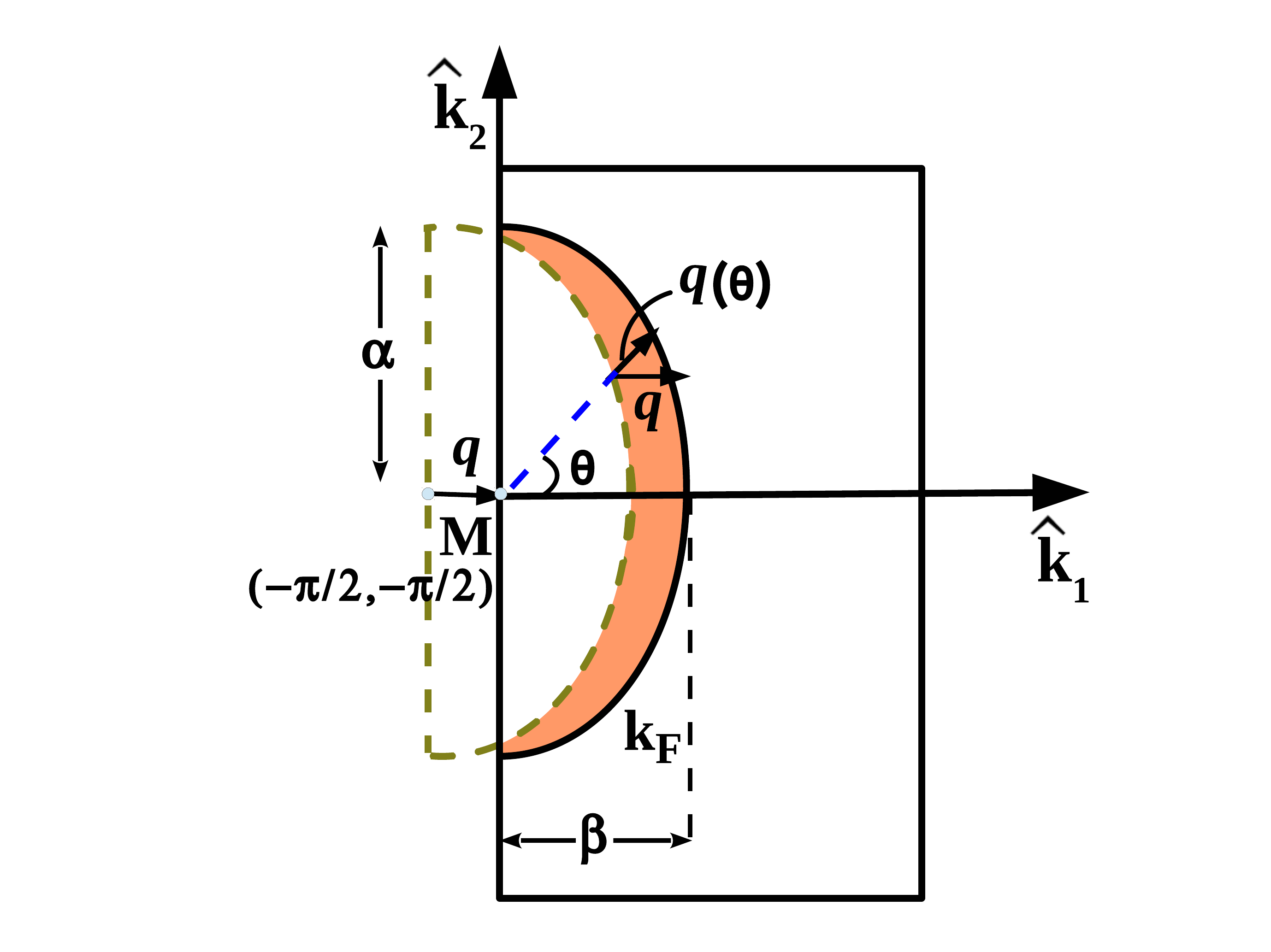}}
  \caption{ (color online) Figure used to evaluate the integral in Eq. (\ref{lamdamax2}) for the  intra-band contribution to the transverse response.
Only the electron states  within the crescent-shaped shaded region  contribute to the  integral. 
We have taken the $q$ vector along $M \rightarrow \Gamma$ direction, $\hat k_1 = (\hat k_x + \hat k_y) / \sqrt 2 $ for simplicity of integration, since the result is only weakly dependent on the direction of $q$
as discussed in the text.
  }
\label{fig4}
\end{figure} 

This summation can be easily performed for small $q$ ($q \ll \alpha, \beta$), where $\alpha$ and $\beta$ are respectively the semi-major and semi-minor axes of the Fermi ellipse.  
Converting the summation into integration, and noting that only the crescent-shaped region in Fig. \ref{fig4} contributes to the integral, Eq. (\ref{lambdamax1}) becomes 
\begin{eqnarray}      \label{lamdamax2}
[\lambda^{\rm max}_{\perp}(q)]^{\rm intra} = \frac{2}  {(2\pi)^2} 
\int_{-\pi/2 }^{ \pi / 2} d\theta                     \int_{k_F (\theta)-q (\theta)}^{k_F (\theta)}    k \ dk \times \nonumber \\
\hspace{10mm}
\frac{ U (\varepsilon^{(1)}_k-\varepsilon^{(1)}_{k+q})^2/   (4\Delta^2) }
{(\varepsilon^{(2)}_{k+q}-\varepsilon^{(2)}_k+\varepsilon^{(3)}_{k+q}-\varepsilon^{(3)}_k)+((\varepsilon^{(1)}_{k+q})^2-(\varepsilon^{(1)}_k)^2)/  (2\Delta) },
\end{eqnarray}
where a factor of two is included in the numerator to take into account the contributions from the two spin bands as indicated in Fig. \ref{fig3}. 
The crescent region in the integral is defined by the radial vector going between $k_F (\theta) =  \beta \sec \theta /(1+\beta^2 \tan^2 \theta/ \alpha^2)^{1/2}$
and $k_F (\theta)- q (\theta)$, where  
$q(\theta) = q\sec \theta /(1+\beta^2 \tan^2 \theta/ \alpha^2)$. 
The spiral wave vector $q$ is taken along the $\hat k_1$ direction here, which is justified, since  as discussed later, the result is only weakly dependent
on the direction of $q$.
Note also that the lower limit of the $k$ integral in Eq. (\ref{lamdamax2}) does have an $O \ (q^2)$ term, which however would contribute to the result only in higher order in $q$.

The integral (\ref {lamdamax2}) can be performed in a straightforward manner using the Taylor  expansion,
$\varepsilon_{k+q}-\varepsilon_k = q \cdot   [\nabla{\varepsilon_k} ]_{k = k_F (\theta)}  $. 
The result is 
\begin{eqnarray}\label{lambda-intra}
[\lambda^{\rm max}_{\perp}(q)]^{\rm intra} = \frac{ U t_1^2 q^2}{4\Delta^3}  \times  \alpha_{\rm intra},      \nonumber \\
 \alpha_{\rm intra} = \Delta (2t_3+t_2)^{-1} \times [1/(2\pi\eta) -x/(4\eta^2) -x/4] ,
\end{eqnarray}
where the relation $\alpha\beta = \pi x$, connecting the semi-major and semi-minor axes of the Fermi ellipse to the concentration of the doped electrons $x$ (per Ir atom) has been used. The constant $\eta \equiv \beta / \alpha$ is related to the eccentricity $e$ of the  ellipse according to the relation $e  = \sqrt{1-\eta^2}$. 
It turns out that for $q$ along $\hat k_1$ considered here, Eq.  (\ref {lambda-intra}) is the total contribution to the lowest  order in $x$. It is easily seen by inspection that the rightmost half-ellipse in Fig. (\ref{fig2} b) contributes nothing as there are no final states that the electrons can go to, while 
explicit calculation shows that the top and the bottom half-ellipses in the same figure contribute only to the order $O \ (x^2)$.

{\it Angle dependence of the spiral wave energy} --
 In the above calculation, we took the spiral wave vector $q$ along $\hat k_1$ for the ease of calculation, which is justified because the energy 
 of the spiral state depends only weakly on the direction of the $q$ vector, and therefore the instability criterion can be obtained by taking $q$ along any direction without making significant error. 
  
\begin{figure}[h]
 \centerline{\includegraphics[width=250pt]{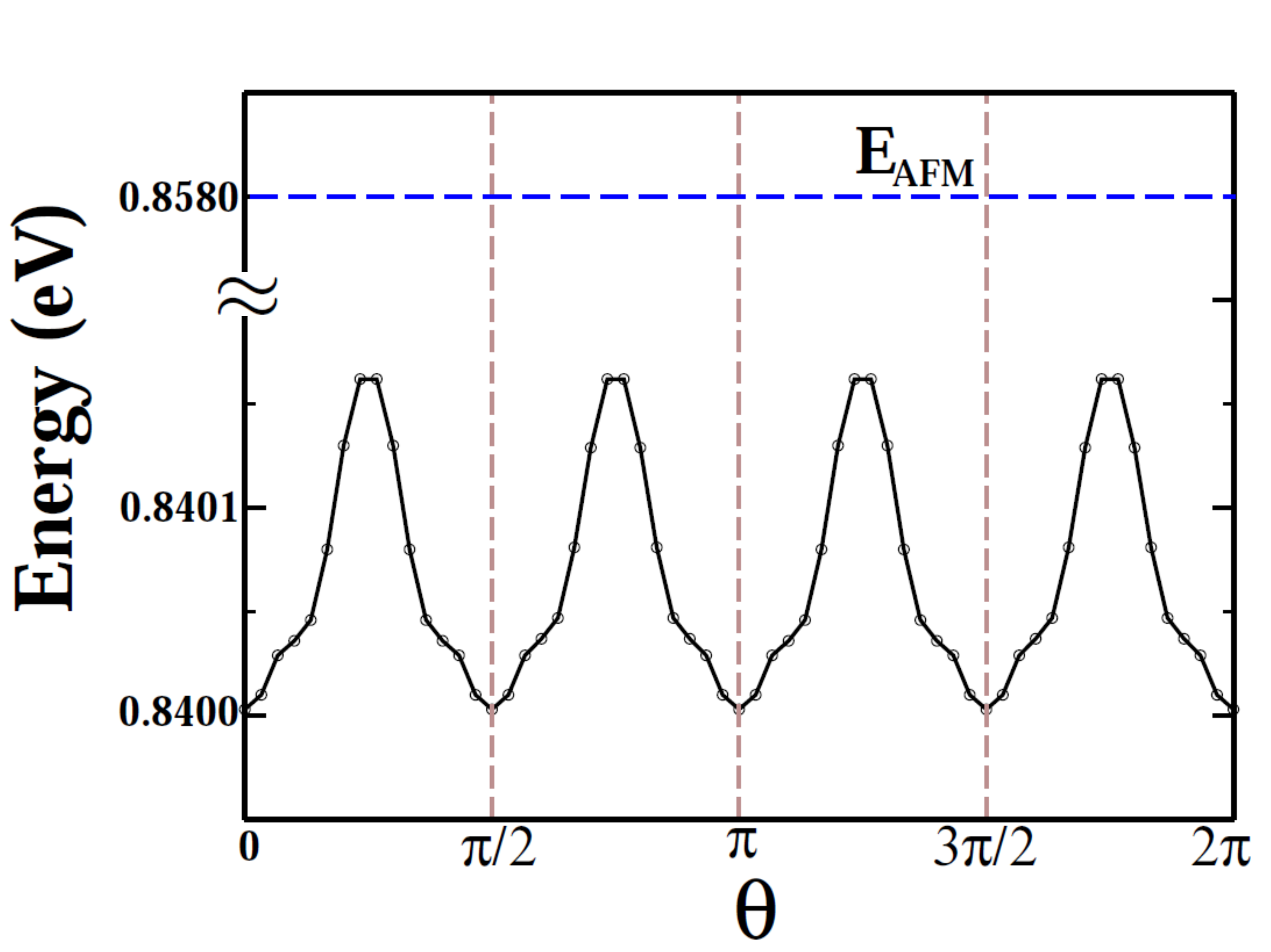}}
  \caption{ (color online)  Variation of the total energy of the Hubbard model for the spiral state with the wave vector $ q = \pi / 10 \ (\cos \theta, \sin \theta)$ , i. e., along the direction $\theta$ with respect to $\hat k_x$ in the $k_x - k_y$ plane,
  for a case where the AFM state is unstable. The hopping parameters $t_i$ are the same as in Fig. \ref{fig4}, $ U = 3$ eV, and doping concentration $x = 0.3$. The total energy of the AFM state is also shown in the figure indicating that it  is unstable. The ground state
  (not indicated in the figure)
  corresponds to a spiral state with a different $q$, and has a lower energy (0.836 eV) than the case shown here, as it must.
  Note that the angle dependence of the energy for the spiral state, which arises due to the ellipticity of the Fermi surface, is very small as compared to the energy difference from the AFM state. 
  }
\label{fig7}
\end{figure}

To see this, we compute the energy of the spiral wave state for the Hubbard Hamiltonian (\ref{Hubbardmodel}) following methods used in our earlier work \cite{Jamshid}.
The variation of the total energy of the spiral state as a function of the direction $\theta$ of the spiral vector is shown in Fig. \ref{fig7} for a typical case. As we can see from the figure, the direction dependence of the spiral state energy is exceedingly weak. 
The energy can be roughly described by the expression $ E (q) = E_{AF} + Aq^2 + Bq^2 \sin^2 2\theta$, with $A \approx -0.2$ eV
 and  $B \approx 2\times10^{-3}$ eV, so that the angle dependence is weak.  
 This justifies our choice of the direction of $q$ in the calculation of the intra-band contribution,
 which was chosen in order to make the integrations simple. 
 
\subsection{Inter-band Contribution}   

\begin{figure}[h] 
  \centerline{\includegraphics[width=200pt]{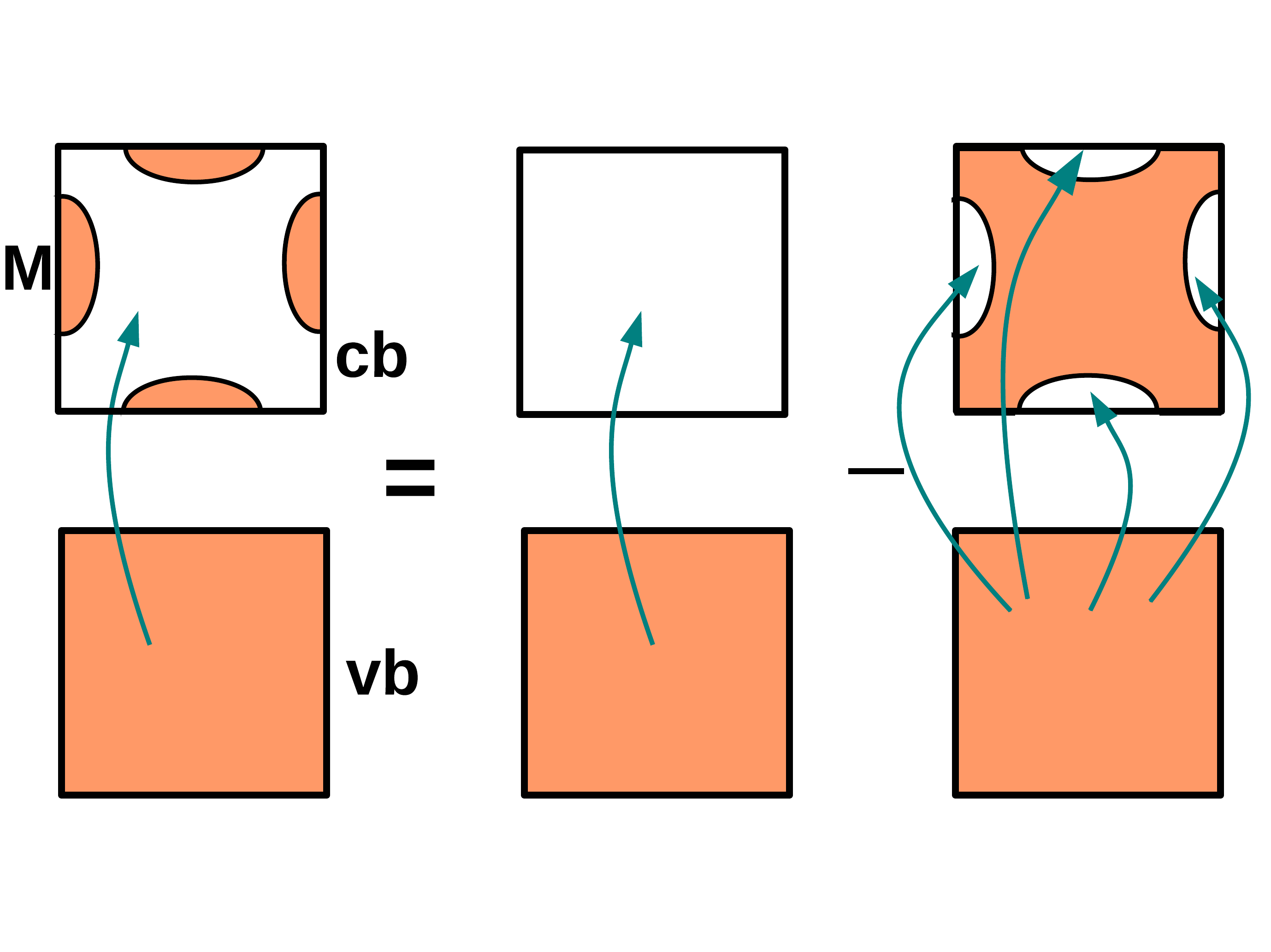}} 
  \caption{ (Color online) Splitting the summation  in the $k$-space  for the evaluation of   Eq.  (\ref{interband})
  into two parts, $I_1$ and $I_2$, given in Eq.  \ref{chiABinter}. The squares represent the Brillouin zones and the occupied valence and conduction states are indicated by colored areas.
  }
\label{figintegral}
\end{figure}

We now turn to the inter-band contribution. 
The inter-band processes involve the virtual transitions of electrons from the completely filled valence band
 to the empty states in the conduction band as indicated in Fig. \ref{figintegral}. 
 We first compute the off-diagonal element
 of $\chi$. From Eq. (\ref{instability}) and Eqs. (\ref{eigenval}) and (\ref{ef2}), we have 
\begin{eqnarray} \label{interband}
[ \chi_{AB}{(q)}]^{\rm inter} = U
 \sum_k^{\prime}     \left( \frac{\psi^{v}_{k\uparrow}\psi^{c}_{k+q\downarrow}\phi^{c} _{k+q\downarrow}\phi^{v} _{k\uparrow}}{\varepsilon^{c}_{k+q\downarrow}-\varepsilon^{v}_{k\uparrow}}
+ \frac{\psi^{v} _{k\downarrow}\psi^{c} _{k+q\uparrow}\phi^{c} _{k+q\uparrow}\phi^{v} _{k\downarrow}}{\varepsilon^{c}_{k+q\uparrow}-\varepsilon^{v}_{k\downarrow}}\right) \nonumber \\
\hspace{10mm}
 = - U\sum^{\prime}_k \varepsilon^{(1)}_k\varepsilon^{(1)}_{k+q}/(4\Delta^3),                    
\end{eqnarray}    
where we have kept only terms $ O \ (1/\Delta^3)$ and the prime over the summation, again, means that the summation
goes over all occupied states of momentum $k$ in the valence band (here the full Brillouin zone) 
to the unoccupied states of momentum $k+q$ in the conduction band. 
Unlike the case of the intra-band contribution, where we took the spiral wave vector $q$ along a specific direction $\hat k_1$ for the ease of calculation, for the inter-band contribution, any direction of $q \equiv (q_x, q_y)$   can be treated equally easily, which we treat below.

The summation (\ref{interband}) is conveniently evaluated by splitting it into two parts as indicated in Fig. \ref{figintegral} and taking the difference. Converting the summation into integration, we have the results for the two parts  

%
\begin{eqnarray} \label{chiABinter}
&& [ \chi_{AB}{(q)}]^{\rm inter} = I_1 - I_2 , \nonumber \\
&& I_1 = 
\frac{-2U}{(2 \pi)^2}      \int_{0}^{\sqrt{2}\pi}     dk_1 \int_{0}^{\sqrt{2}\pi} dk_2~\varepsilon^{(1)}_k\varepsilon^{(1)}_{k+q}/(4\Delta^3)  
= -Ut_1^2\gamma_q/\Delta^3, \nonumber \\
&& I_2 =  4 \times
\frac{-2U}{(2\pi)^2}   \int_{-\frac{\pi}{2}} ^{\frac{\pi}{2}}   d\theta \int_0^{k_F (\theta)}   k dk~\varepsilon^{(1)}_k\varepsilon^{(1)}_{k+q}/ (4\Delta^3)  \nonumber \\
&&  = -t_1^2(\sin q_x + \sin q_y)\beta^3/(\sqrt{2}\pi^2\Delta^3) + O (\beta^4),
\end{eqnarray}
where $\hat k_1$ and $\hat k_2$ are the same directions as in Fig. \ref{fig2}, and $\gamma_q = (\cos q_x + \cos q_y)/2 $. 
The factor of two in the numerator of $I_1$ and $I_2$ is due to the fact that there are two bands for each momentum point in the BZ (corresponding to two Ir atoms in the unit cell). Furthermore,
the factor of four in $I_2$ comes from the four $M$ points in Fig. \ref{figintegral}. Note that $I_2$ is $O (\beta^3)$ 
 and hence neglected as compared to $I_1$.
Similarly, we can calculate the inter-band contribution to the diagonal element 
\begin{eqnarray} \label{chiAAinter}
[ \chi_{AA}{(q)}]^{\rm inter} &=&
  U  \sum_k^{\prime}\left( \frac{(\psi^{v}_{k\uparrow})^2(\psi^{c} _{k+q\downarrow})^2}{\varepsilon^{c}_{k+q\downarrow}-\varepsilon^{v}_{k\uparrow}}
+ \frac{(\psi^{v} _{k\downarrow})^2(\psi^{c} _{k+q\uparrow})^2}  
{\varepsilon^{c}_{k+q\uparrow}-\varepsilon^{c}_{k\downarrow}}\right)  
\nonumber \\
&=& \frac {U} {2 \Delta}
 \sum_k^{\prime} \left( 1 -\frac{\varepsilon (k, q)} {2 \Delta}+\frac{\varepsilon (k, q)^2}{4\Delta^2}-\frac{(\varepsilon^{(1)}_{k+q})^2+(\varepsilon^{(1)}_k)^2}{2\Delta^2} \right),
\end{eqnarray}
where $\varepsilon (k, q) \equiv  \varepsilon^{(2)}_{k+q}-\varepsilon^{(2)}_k+\varepsilon^{(3)}_{k+q}-\varepsilon^{(3)}_k$.
Following the same procedure as described above to perform the integral, we get, after some algebra, the result
\begin {eqnarray}\label{chiAAinter}
[ \chi_{AA}{(q)}]^{\rm inter} &=& U [1/U - 2t_1^2/\Delta^3 + q^2t_2^2/ (2\Delta^3) - t_3^2q^2/\Delta^3 ] \nonumber \\
&-& 4  U [x/(8\Delta)   -   t_1^2q^2x/   (4\Delta^3)].
\end{eqnarray} 
Using Eqs. (\ref{chiABinter}) and (\ref{chiAAinter}) and neglecting higher order terms in $x$, we get the following result for the  inter-band contribution to the maximum eigenvalue
\begin{equation} \label{lambda-inter}
[ \lambda^{-}_\perp(q)]^{\rm inter} = 
[ \chi{(q)}]^{\rm inter}_{AA}-[ \chi{(q)}]^{\rm inter}_{AB}
= 1 -   \frac {U  q^2t_1^2}    { 4\Delta^3} \times \alpha_{\rm inter},
\end{equation}
where 
$\alpha_{\rm inter} = 1- 2t_2^2/t_1^2 + 4t_3^2/t_1^2 - 8x$. 
Adding the intra- and the inter-band contributions, Eqs. (\ref{lambda-intra}) and (\ref{lambda-inter}),  the maximum eigenvalue becomes
\begin{equation} 
\lambda^{\rm max}_\perp(q) = 1 +  \frac{ U q^2t_1^2} {4\Delta^3}     \times (\alpha_{\rm intra}-\alpha_{\rm inter}).
\end{equation}
For the maximum eigenvalue to be less than one, we must therefore have $\alpha_{\rm intra} <  \alpha_{\rm inter} $,  in which case the AFM state would be stable. From Eqs. (\ref{lambda-intra}) and ({\ref{lambda-inter}}), and with the expression of the axis ratio for the 
elliptical Fermi surface $\eta \equiv \beta / \alpha = [ (2 t_3 - t_2) /  (t_2 + 2 t_3 + 4 t_1^2 / U)   ]^ {1/2}$ obtained before, 
the condition that the AFM state is stable reads
\begin{equation}\label{Instability_condition}
8\pi \eta^2 (2 t_3 + t_2) (t_1^2-2t_2^2+4t_3^2-8xt_1^{2})-t_1^2 U (2\eta (1-x)-\pi x-\pi\eta^2 x) > \ 0,
\end{equation}  
where the eccentricity of the Fermi surface is $e = (1 - \eta^2)^{1/2}$.

This is the central result of this work that provides the condition for the stability of the AFM state,
 in terms of the doping concentration $x$ (assumed to be small) and the parameters of the Hubbard model.
 Note that our analysis is valid only if the electron pocket occurs at the $M$ point, which is the case for the iridate. As discussed earlier, in general this happens if condition Eq.  (\ref{condition}) is satisfied.

%
\begin{figure}[t] 
  \centerline{\includegraphics[width=300pt]{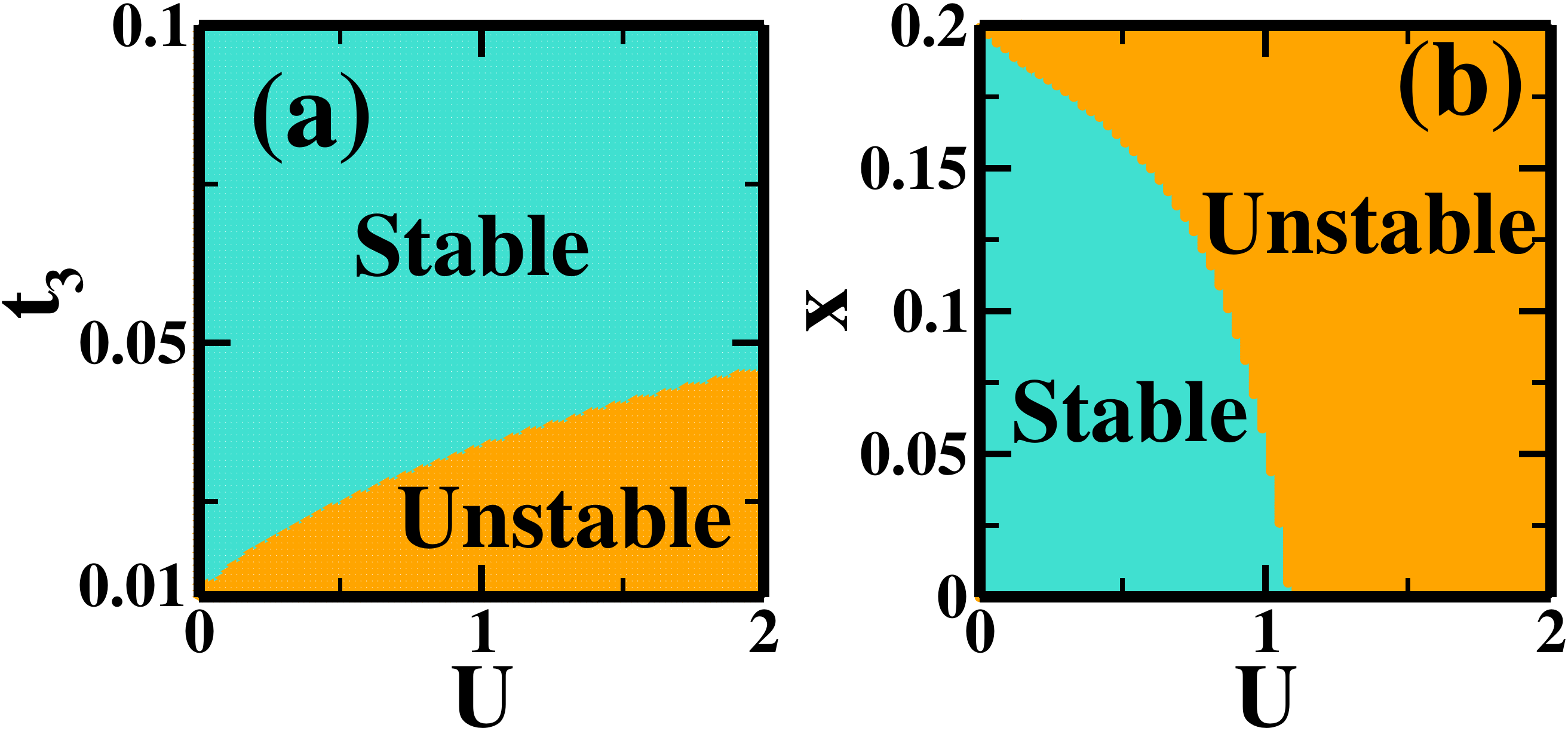}}
  \caption{ (color online) Regions of stability of the AFM phase as obtained from the stability condition Eq.  (\ref{Instability_condition}). 
   Fig. (a) shows the effect of $t_3$ on the stabilization of  the AFM state for the
carrier concentration $x \rightarrow 0$, with all other parameters fixed.  
As per Eq.  (\ref{condition}), we must have $t_3 > t_2 /2 \approx 0.01$, for which results are shown in (a). 
If $t_3 < 0.01$, then the electron pocket shifts away from the $M$ point.
Fig. (b) shows that the AFM state is destabilized beyond a critical value of $x$. 
Parameters used are the same as in Fig. \ref{fig2} unless indicated otherwise in the figures.
}
\label{fig5}
\end{figure} 

\begin{figure}[t]
  \centerline{\includegraphics[width=250pt]{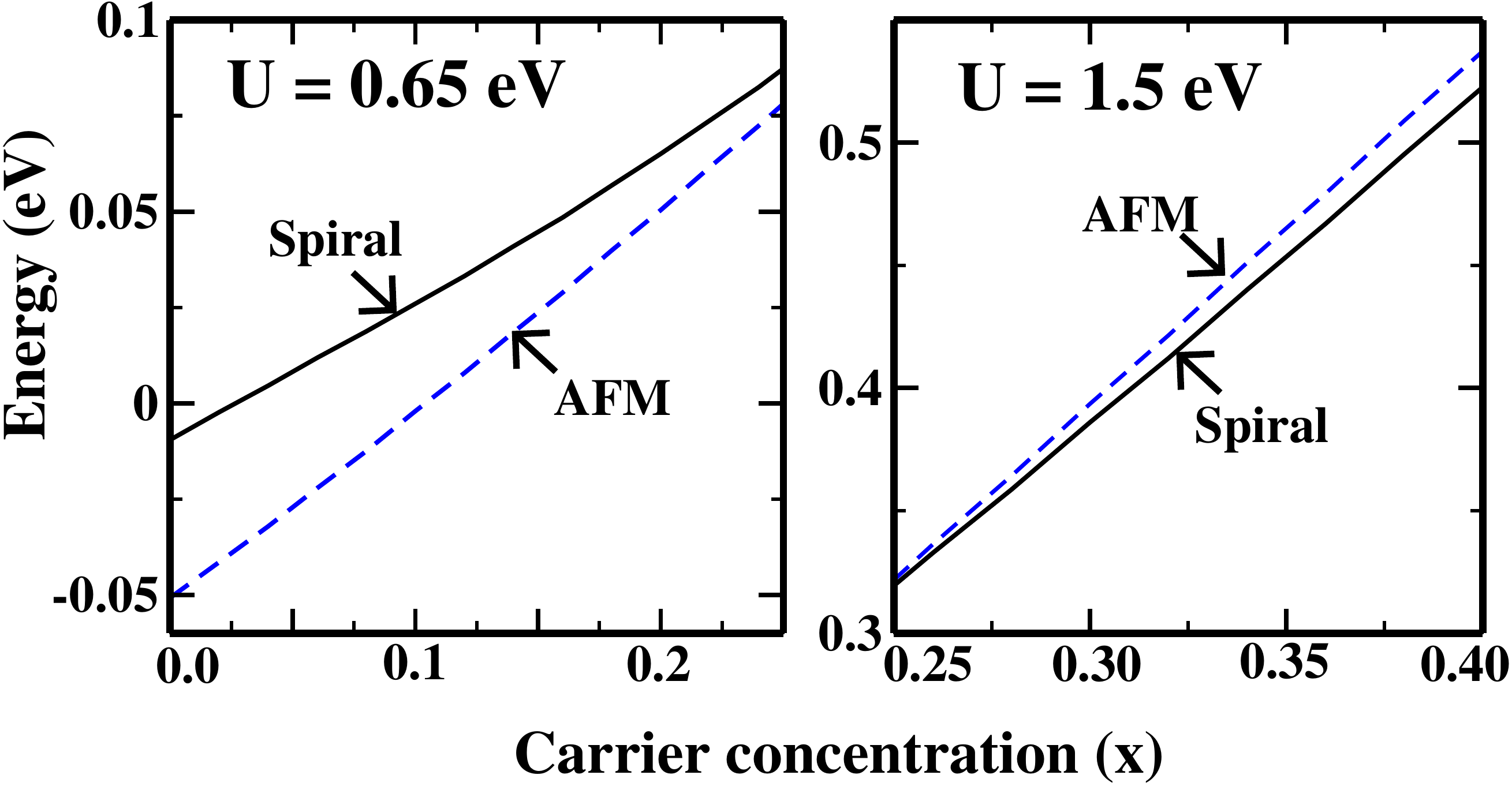}}
  \caption{ (color online)  Energy of the AFM state compared to the  spiral state with the spiral vector, viz., $ q = \pi / 10 \ (1, 1)$, in two regions of stability in Fig. \ref{fig5} (b). Hopping integrals are the same as in  Fig. \ref{fig2}. The right panel is sufficient to show that the AFM state is unstable for $ U = 1.5$ eV for all values of $x$ considered. 
  For the left panel ($ U = 0.65 $ eV), the specific spiral state has always the higher energy; However, some other spiral state (in fact the paramagnetic state, not shown here) has a lower energy than the AFM state for $x \geq  0.15$, 
  making the AFM state unstable, as indicated from the stability regions of Fig. \ref{fig5} (b).
}
\label{fig6}
\end{figure}

The phase diagrams obtained from the instability condition (\ref{Instability_condition})   are shown in Fig. \ref{fig5} for parameters appropriate 
for SIO. 
The stable and unstable AFM regions of the phase diagram in Fig \ref{fig5} (b) are further verified by comparing the total energies of the AFM and the spiral states (Fig. \ref{fig6}), obtained for the Hubbard Hamiltonian (\ref{Hubbardmodel}) following methods developed in earlier works \cite{Jamshid,Arrigoni}.   
As discussed already, a small $t_3$ ($ t_3 > t_2/2$) is necessary for forming the electron pocket at $M$ and it is also necessary to
stabilize the AFM state for the lightly doped system ($x \rightarrow 0$) as seen from Fig. \ref{fig5} (a). 
For $U = 0.65$ eV, $t_3$ must be larger than $\sim 0.025$ eV in order that the AFM state is stabilized.
Fig. \ref{fig5} (b) shows the regions of stability for SIO. We see that for $ U \approx 0.65 $ eV appropriate for SIO, the system remains AFM up to a critical doping concentration of $x_c \approx 15\%$, which is higher than the experimental value of 
$\approx 8\%$ measured in the 
 recent experiments of Chen et al. \cite{Chen},
 but understandable given the simplicity of our one-band model.  
 
 {\it Relation to the Nagaoka Theorem} -- In a seminal paper \cite{Nagaoka}, Nagaoka proved rigorously  for certain lattices that addition of a single carrier (electron or hole) destroys the AFM ground state of the half-filled NN Hubbard model in the strong-coupling limit, turning the system into a FM metal.
Roughly speaking, this happens due to a competition between the kinetic energy $\sim x W$ gained by the carriers by hopping in the FM lattice and the loss of the exchange interaction $\sim W^2 / U$ that favors the AFM lattice, 
leading 
 to the Nagaoka condition for the stability of the AFM phase \cite{Mattis}, viz., $x < \alpha W/U$, valid in the strong-coupling limit and where $\alpha$ is some positive constant. 
 Things are expected to be different when hopping beyond the 1NN are included as is the case in the present problem. For instance, the presence of a strong 2NN hopping would favor the AFM state as the doped carriers can still gain kinetic energy via 2NN hopping (see Fig. \ref{fig1}). When all hoppings are included, Fig. \ref{fig5} shows the regions of stability for the AFM phase. However, since the 2NN and the 3NN hoppings, $t_2$ and $t_3$, are comparatively small as compared to the 1NN hopping $t_1$, Fig. \ref{fig5} (b) is consistent with the Nagaoka condition in the large $U$ limit, viz., that infinitesimal doping destabilizes the AFM state. For $U \approx 0.65$ eV appropriate for SIO, a larger carrier concentration $x_c\approx 15\%$ is needed to destabilize the same state.

{\it Stability of the AFM phase in the undoped system ($ x = 0$)} --
To be consistent, we must show that the AFM state in SIO is stable for the half-filled case, {\it i. e.}, with no doped electrons.
The instability condition (\ref{Instability_condition}) was obtained for small doping (small $x$ and therefore small $k_F$) and small spiral wave vector $q$, with the condition that $ |q| \ll k_F$. We still want to study the instability for the half-filled case, where $x = k_F = 0$, 
and $|q| $ is small but non-zero. Therefore, we can't use Eq.  (\ref{Instability_condition}) for the stability. However, in this case we just have the inter-band transitions contributing to the response matrix $\chi (q)$, and referring to Fig. \ref{figintegral}, only the first term on the right hand side contributes, which we have already evaluated. The $\chi_{AB}$ and $\chi_{AA}$ are, respectively, $I_1$ in Eq.  (\ref{chiABinter}) and
the first term in Eq.  (\ref{chiAAinter}). The maximum eigenvalue thus becomes, for $x = 0$,
\begin{equation} 
\lambda^{\rm max}_\perp(q) = 1 - \frac{U}{2 \Delta^3} \times   (2t_1^2 -q^2 t_2^2 + 2q^2 t_3^2).
\end{equation}
Clearly, for small $q$, $\lambda^{\rm max}_\perp(q) < 1$, indicating that the AFM phase is stable with respect to the spiral state.

{\it Stability with respect to the formation of charge-density waves} --
In the above discussions, we studied the instability of the AFM state with respect to the formation of the spiral SDW states. On general grounds, one expects the charge density wave (CDW) states to be energetically unfavorable, unless special situations exist such as the Fermi surface ``nesting." In the strong-coupling limit ($U/W \rightarrow \infty$), where band structure energy
can be neglected, it is easy to see that the Coulomb interaction would prefer an equal charge distribution on all sites in the lattice,
so that the system is stable with respect to the formation of any CDW.  
In contrast, the SDW state does not require any charge imbalance between sites, so that the system is more prone to the SDW instability as was seen above. 

Since the strong-coupling limit is reasonably satisfied  and the there is no Fermi surface ``nesting" (Fig. \ref{fig2}), CDW instability is less likely for SIO. Mathematically, this can be shown by computing the response function for the perturbing longitudinal field
$\Delta^i =  U  \langle c^\dagger_{i\uparrow} c_{i\uparrow} + c^\dagger_{i\downarrow} c_{i\downarrow} \rangle$, which
will change the charge amount at the $i$-th site. This longitudinal field is added to the diagonal elements of the Hartree-Fock Hamiltonian.
As before, we consider the instability to the CDW, with the wave vector $q$
with the transverse field $ (f_A, f_B) \exp \ (i   q \cdot   R_i)$,
where  
  $f_A= U \langle c^\dagger_{A\uparrow} c_{A\uparrow} + c^\dagger_{A\downarrow} c_{A\downarrow} \rangle$ and $f_B= U \langle c^\dagger_{B\uparrow} c_{B\uparrow} + c^\dagger_{B\downarrow} c_{B\downarrow} \rangle$ 
  are the order-parameter amplitudes on the two sublattice atoms in the unit cell. 
  After some tedious but straightforward algebra
  following the procedure of the previous sections, we find the result for the maximum eigenvalue of the response matrix to be
\begin{equation} 
\lambda^{\rm max}_{||} (q) = -\frac{t_1^2U}{2 \Delta^3} [1-\frac{\Delta x}{4 (t_2 + t_3)} ]+ O (q),
\end{equation}
which is less than one,
indicating that the AFM state is stable with respect to the formation of any CDW.
In fact, $\lambda^{\rm max}_{||}$ tends to zero in the strong -coupling limit, implying that any CDW fluctuation is very quickly renormalized to zero, as the applied fluctuation $f$ leads to the resultant fluctuation $\lambda^{\rm max}_{||} f$ in the next cycle.

\section{SUMMARY} \label{sec4}

In summary, we examined the stability of the AFM ground state in the strong spin-orbit coupled material Sr$_2$IrO$_4$ with electron doping. The system is of considerable interest owing to the spin-orbital entangled nature of the doped electrons, which occupy
the upper Hubbard band of the Mott insulating state.

Our starting point was a multi-neighbor one-band Hubbard model that describes the basic features of the J$_{\rm eff} = 1/2$ sector of the 
band structure. It is a one-band model, although at least three NN interactions must be included
in order to describe the main features of the band structure. 
The parameters in the Hubbard model can be varied, but certain conditions [specifically, Eq.  (\ref{condition})] apply
in order that the electrons occupy the $M$ pocket in the Brillouin zone as predicted from the density-functional calculations.

We studied the stability of the AFM ground state  to the formation of a spiral SDW state in the strong coupling limit  by applying a perturbing transverse field with a spiral wave vector and examining if the transverse
field is enhanced or diminished by computing the transverse response using linear response within the Hartree-Fock mean-field theory. 
Note that the paramagnetic and ferromagnetic states are special cases of the spiral state and are therefore included in the stability analysis.
The condition for the stability of the AFM phase was obtained in terms of the parameters of the Hubbard model and the eccentricity of the Fermi surface for small dopant concentration.
The resulting stability regions were indicated as a phase diagram in the $ x - U$ space (Fig. \ref{fig5}).
The results indicate that the higher neighbor interactions, especially the relatively strong 3NN hopping, 
play an important role in the stability of the AFM state. For parameters appropriate for Sr$_2$IrO$_4$, the AFM region extends up to the critical  doping concentration $x_c$ ($0 < x < x_c$), where $x_c \sim 0.15$.
This qualitatively explains the anti-ferromagnetism (long or short range) observed in the experiments for the electron doped system with $x_c \approx 0.08$.

Finally, we note that our results are not immediately transferrable to the hole-doped system, since there is no electron-hole symmetry in the problem -- indeed, in this case, the hole pocket occurs at a different point in the Brillouin zone. However, it is straightforward to extend the stability analysis to the hole doped system as well.
  
\ack
We thank Avinash Singh for stimulating discussions and the U.S. Department of Energy, Office of Basic Energy Sciences, Division of Materials Sciences and Engineering for financial support under Grant No. DEFG02-00ER45818.


\nocite{*}
\bibliographystyle{iopart-num}%
\bibliography{JPC}%

\end{document}